\documentclass{article}
\usepackage{fortschritte}
\usepackage{amsmath, amssymb}
\usepackage{graphicx}
\usepackage{dcolumn}
\usepackage{bm}
\usepackage{amssymb}
\def\beq{\begin{equation}}                     %
\def\eeq{\end{equation}}                       %
\def\bea{\begin{eqnarray}}                     
\def\eea{\end{eqnarray}}                       
\newwrite\ffile\global\newcount\figno \global\figno=1

\def\writedef#1{}

\input epsf
\def\figin{\epsfcheck\figin}\def\figins{\epsfcheck\figins}
\def\epsfcheck{\ifx\epsfbox\UnDeFiNeD
\message{(NO epsf.tex, FIGURES WILL BE IGNORED)}
\gdef\figin##1{\vskip2in}\gdef\figins##1{\hskip.5in}
\else\message{(FIGURES WILL BE INCLUDED)}%
\gdef\figin##1{##1}\gdef\figins##1{##1}\fi}

\def\figinsert{}
\def\ifig#1#2#3{\xdef#1{fig.~\the\figno}
\writedef{#1\leftbracket fig.\noexpand~\the\figno}%
\figinsert\figin{\centerline{#3}}\medskip\centerline{\vbox{\baselineskip12pt
\advance\hsize by -1truein\center\footnotesize{  Fig.~\the\figno.} #2}}
\bigskip\endinsert\global\advance\figno by1}
\def\endinsert{}

\def\N{{\cal N}}
\begin {document}                 
\def\titleline{
%
%
%
A Gravity Dual of Chiral Symmetry Breaking    
}
\def\email_speaker{
{\tt 
%
%
ik@physik.hu-berlin.de                       
}}
\def\authors{
%
%
%
%
%
J.~Babington\1ad, J.~Erdmenger\1ad,          
N.~Evans\2ad, Z.~Guralnik\1ad,               
I.~Kirsch\1ad\sp                             
}
\def\addresses{
%
%
%
%
\1ad                                        %
Institut f\"ur Physik, Humboldt-Universit\"at zu Berlin\\ %
Newtonstra\ss e 15, D-12489 Berlin, Germany\\ 
\2ad                                        
Department of Physics, Southampton University\\    
Southampton  S017 1BJ,
United Kingdom             
   %
}
\def\abstracttext{
%
%
  We establish a holographic description of chiral symmetry breaking for a
  particular confining large $N$ non-supersymmetric gauge theory with matter
  in the fundamental representation. The gravity dual of this theory is
  obtained by introducing a D7-brane probe into a deformed Anti-de~Sitter
  background found by Constable and \mbox{Myers}.  We numerically compute the
  $\bar\psi\psi$ condensate and meson spectrum by solving the equation of
  motion of the D7-brane probe in this background.  We find a symmetry
  breaking condensate as well as the associated, pion-like, Goldstone boson in
  the limit of small quark mass.  The existence of the condensate ensures that
  the D7-brane never reaches the naked singularity at the origin of the
  deformed AdS space.
}
\large
\makefront
\section{Introduction}
Generalizations of AdS/CFT duality  \cite{maldacena} in which
supersymmetry and conformal symmetry are broken have led to
promising new ideas for studying strong coupling phenomena in
large $N$ gauge theories.  In particular, it is hoped that methods
based on gauge-gravity duality will eventually be applicable to
QCD.  Notable progress towards this goal has been achieved by
considering asymptotically AdS geometries which correspond to renormalization
group (RG) flows from a super-conformal gauge theory in the
ultraviolet to a QCD-like theory in the infrared. A number of
non-supersymmetric ten-dimensional geometries of this form have
been found
\cite{Witten,Gubser:1999pk,Babington:2002ci, CM}
and been shown to describe confining gauge dynamics. There have
been many interesting calculations of the glueball spectrum
in three- and four-dimensional QCD-like theories 
in such backgrounds (see for instance~\cite{Csaki}).
These spectra are obtained by solving the
classical supergravity equations of motion. While these geometries are
actually strong-coupling descriptions of
gauge theories with extra massive degrees of freedom in
addition to those of pure QCD,  the glueball spectra still compare
favorably with lattice data \cite{Teper}.

To take a step closer to finding a holographic dual of QCD,  it is
necessary to include particles in the fundamental representation
of the gauge group, i.e.~quarks. 
The addition of fundamental degrees of freedom to 
the dual gauge theory is achieved by embedding an appropriate D-brane 
probe in the gravity background. Holography in the presence of a probe
brane was studied in detail in relation to defect conformal field
theories \cite{KarchRandall,DFO,CEGK1}.

To obtain a holographic dual of a four-dimensional gauge theory with
fundamental matter,  Karch and Katz \cite{KarchKatz} considered a 
configuration in which a D7-brane in $AdS_5 \times S^5$ fills $AdS_5$
and wraps an $S^3$ inside $S^5$. This configuration is the
near-horizon limit of $N$ D3-branes sharing three spatial directions
with a single D7-brane. The open string degrees of freedom are those of the
${\mathcal N} =4$ super Yang-Mills theory, coupled to an
${\mathcal N} =2$ hypermultiplet in the fundamental
representation of $SU(N)$. The latter arises from strings stretched
between the D7- and D3-branes.  When the D7 and D3's are
separated, the fundamental matter becomes massive  and the dual
description of the probe D7-brane has an induced metric that is
only asymptotically $AdS_5 \times S^3$.  In this case there is a
discrete spectrum of mesons. This spectrum has been computed
(exactly!)\ at large 't~Hooft coupling \cite{MateosMyers} using an
approach analogous to the glueball calculations in deformed AdS
backgrounds. The novel feature here is that the ``quark'' bound
states are described by the scalar fields in the Dirac-Born-Infeld
(DBI) action of the D7-brane probe.  Neglecting the back-reaction
of the probe on the geometry corresponds to the quenched
approximation of the gauge theory. -- Further related work on 
embedding a D7 probe into the Klebanov-Strassler background may
be found in \cite{Sakai:2003wu}.

In this talk we consider one of the most important features of QCD dynamics,
which is chiral symmetry breaking by a $\bar \psi \psi$ condensate.  This
occurs only in non-supersymmetric theories. Therefore we consider as in
\cite{Babington:2003vm} the embedding of a D7-brane probe into the
non-super\-symmetric background of Constable and Myers \cite{CM}, which
exhibits confinement.  In this holographic setting we do indeed find chiral
symmetry breaking.

The Constable-Myers background is asymptotically $AdS_5 \times
S^5$, but has a non-constant dilaton and $S^5$ radius. In the dual
gauge theory this corresponds to the addition of an operator of
dimension 4 with zero R-charge. This generates an RG flow from
$\N=4$ super Yang-Mills to a non-supersymmetric gauge theory. 
The gauge theory does not quite
have the field content of QCD, since the adjoint fermions and
scalars of the ultraviolet ${\mathcal N}=4$ theory do not
decouple. Nevertheless, in a
certain parameter range,  the geometry leads to an area law for the
Wilson loop, corresponding to confinement in the dual gauge theory.
Solving the classical gravity equations also gives a discrete spectrum
of glueballs with a mass gap.  

We obtain numerical solutions for the D7-brane equations of motion
in the Constable-Myers background with UV asymptotic behavior
determined by a quark mass and a chiral condensate. The quark mass
$m$ and the quark condensate expectation value $c$ are given by
the UV asymptotic behavior of the solutions to the
Dirac-Born-Infeld equations of motion in the standard holographic
way.\footnote{In the $\N=2$ supersymmetric scenario of
\cite{KarchKatz} with a D7-brane probe in standard AdS space,
there cannot be any regular solution which has $c\neq 0$; the
supersymmetric theory does not allow a quark condensate. We
confirmed this by an explicit calculation.} Using a numerical
`shooting' technique, we compute the condensate $c$ as a function
of the quark mass $m$ by imposing a regularity constraint. We find
that there are regular solutions with non-vanishing chiral
condensate.

A significant feature of the Constable-Myers geometry is that it
has a naked singularity in the far infrared, whose interpretation
is a delicate issue, for instance in the light of the analysis of
\cite{Gubser:2000nd}. Remarkably, our results are not sensitive to
the singular behavior of the metric in the IR. For all the
solutions of the D7 equation of motion satisfying the regularity
constraint, the D7 ``ends'' before reaching the curvature
singularity. Of course the D7-brane does not really end, however
the $S^3$ about which it is wrapped contracts to zero size, in a
manner similar to a scenario discussed in \cite{KarchKatz}. The
screening of the singularity is related to the existence of the
condensate, which persists even in the limit $m\rightarrow 0$.
This corresponds to spontaneous breaking of a $U(1)$ chiral
symmetry of the dual gauge theory. In the probe geometry this
$U(1)$ is the rotation symmetry in the two directions orthogonal
to the D7-brane worldvolume. We emphasize that this $U(1)$ chiral
symmetry is non-anomalous in the large $N$ limit
\cite{Wittenetaprime},  which is also the limit in which the
classical supergravity approximation is valid.

We also compute the meson spectrum by studying classical
fluctuations about the D7-embedding. In agreement with spontaneous
chiral symmetry breaking, the meson spectrum contains a massless
mode in the limit of zero quark mass.

For subsequent developments concerning gauge/gravity duality with 
flavor see Refs.~\cite{Wang:2003yc, Ouyang, Nunez:2003cf, 
Kruczenski:2003uq, Hong:2003jm}. This note is based on a talk given by 
I.~Kirsch at the 36th Symposium Ahrenshoop: Recent Developments in 
String/M-Theory and Field Theory, August 2003.

\section{D7 probe brane in Constable-Myers background}

We choose an appropriate coordinate system for the background of
\cite{CM}
such that in {\it Einstein frame}, the geometry is given by
\begin{align} ds^2 & =  H^{-1/2} \left( { w^4 + b^4 \over w^4-b^4}
\right)^{\delta/4} \, \sum\limits_{j=0}^{3} dx_{j}^2 
+ H^{1/2} \left( {w^4 + b^4 \over w^4-
b^4}\right)^{(2-\delta)/4} {w^4 - b^4 \over w^4 } \sum_{i=1}^6
dw_i^2 \,, \end{align} where $b$ is the parameter of the geometry 
that determines the size of the deformation ($\delta = L^4/(2 b^4)$
with $L$ the AdS radius)  and
\vspace{-0.2cm}
\begin{align} H =  \left(  { w^4
+ b^4 \over w^4 - b^4}\right)^{\delta} - 1 \, ,  \;\;
w^2 = \sum\limits_{i=1}^{6} w_i{}^2
\, .
\end{align}
In this coordinate system, the dilaton and four-form become,
with $\Delta^2 + \delta^2 =10$,
\beq e^{2 \phi} = e^{2 \phi_0} \left( { w^4 + b^4 \over
w^4 - b^4} \right)^{\Delta}, \;\; C_{(4)} = - {1 \over 4}
H^{-1} dt \wedge dx \wedge dy \wedge dz \,.
\eeq
We now consider the D7-brane action in the static gauge with
world-volume coordinates identified with $x_{0,1,2,3}$ and
$w_{1,2,3,4}$,  with transverse fluctuations parameterized by
$w_5$ and $w_6$.  It is convenient to define a coordinate
$\rho$ such that $\sum_{i=1}^4 dw_i^2 = d\rho^2 + \rho^2
d\Omega_3^2$ and the radial coordinate is given by
$w^2=\rho^2 + w_5{}^2 + w_6{}^2$.
The Dirac-Born-Infeld action of the D7-brane probe in
this background takes the form
\begin{align} S_{D7} & = - T_7
\int d^8 \xi~ \epsilon_3 ~  e^{ \phi} { \cal G}(\rho,w_5,w_6)
\nonumber \Big( 1 + g^{ab} g_{55} \partial_a w_5
\partial_b w_5  + g^{ab} g_{66} \partial_a w_6
\partial_b w_6 \Big)^{1/2},
\end{align}
where
\begin{align} & {\cal G}(\rho,w_5,w_6)
 = \rho^3 {( (\rho^2 + w_5^2
+ w_6^2)^2 + b^4) ( (\rho^2 + w_5^2 + w_6^2)^2 - b^4) \over
(\rho^2 + w_5^2 + w_6^2)^4} \,. \nonumber
\end{align}
 From these equations we derive the corresponding equation of
motion. We look for classical solutions of the form $w_6 =
w_6(\rho), w_5 =0$ that define the ground state. Numerically we
find solutions with the asymptotic behavior $w_6 \sim m +
c/\rho^2$. The identification of these constants as field theory
operators requires a coordinate transformation because the scalar
kinetic term is  not of the usual canonical AdS form. Transforming
to coordinates \cite{KarchKatz} in which the kinetic term has
canonical form, we see that $m$ has dimension 1 and $c$ has
dimension 3. These coefficients are then identified with the quark
mass $m_q$ and condensate $\langle \bar{\psi} \psi \rangle$
respectively, in agreement with the usual AdS/CFT dictionary.

We find the physical solutions of the equation of motion by
imposing the following regularity constraint: The
Constable-Myers background has a naked singularity in the far IR at
$\rho^2 + w_6^2 = b^2$ (We set $b=1$ henceforth).
Thus there are  two possibilities for a
solution with an interpretation as an RG flow: either
the D7-brane terminates at a value of $w=w_6(\rho=0)$ away
from the naked singularity via a collapse of the $S^3$, or the
D7-brane reaches the singularity.  In the latter case
we would have little control over the physics without a better
understanding of string theory in such highly curved backgrounds.
This applies in particular to the solution with $m=0$. For
$m=0$, the solution $w_6 =0$ is exact,
which naively seems to indicate the absence of a chiral condensate
$(c=0)$. However, this solution reaches the singularity, and
therefore can not be trusted.

Fortunately something remarkable seems to happen.  For non-zero
values of $m$ we find that the regular solutions terminate before
reaching the singularity.  Some of these regular solutions are
plotted in figure 1a.   Even for a very small but non-zero mass,
the regular solutions obtained numerically require a non-vanishing
$c$ and terminate at $w\geq 1.35$ before reaching the singularity
at $w=1$! This suggests that for $m=0$ there are two solutions. In
addition to the unphysical solution $m=0$, $c=0$ which reaches the
singularity, there is a regular solution $m=0$, $c=1.85$ which is
obtained when taking the limit $m\rightarrow 0$. 
\begin{figure}[!h]
\begin{center}
\includegraphics[height=4cm,clip=true,keepaspectratio=true]{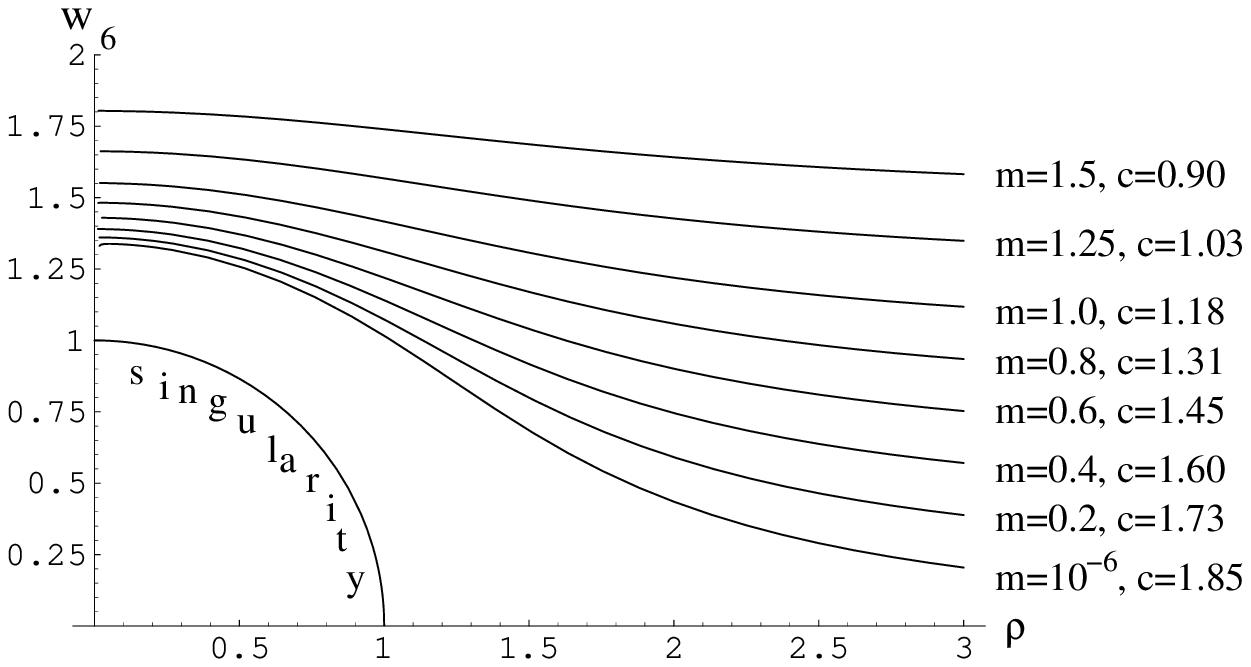}
\hspace{.5cm}
\includegraphics[height=4cm,clip=true,keepaspectratio=true]{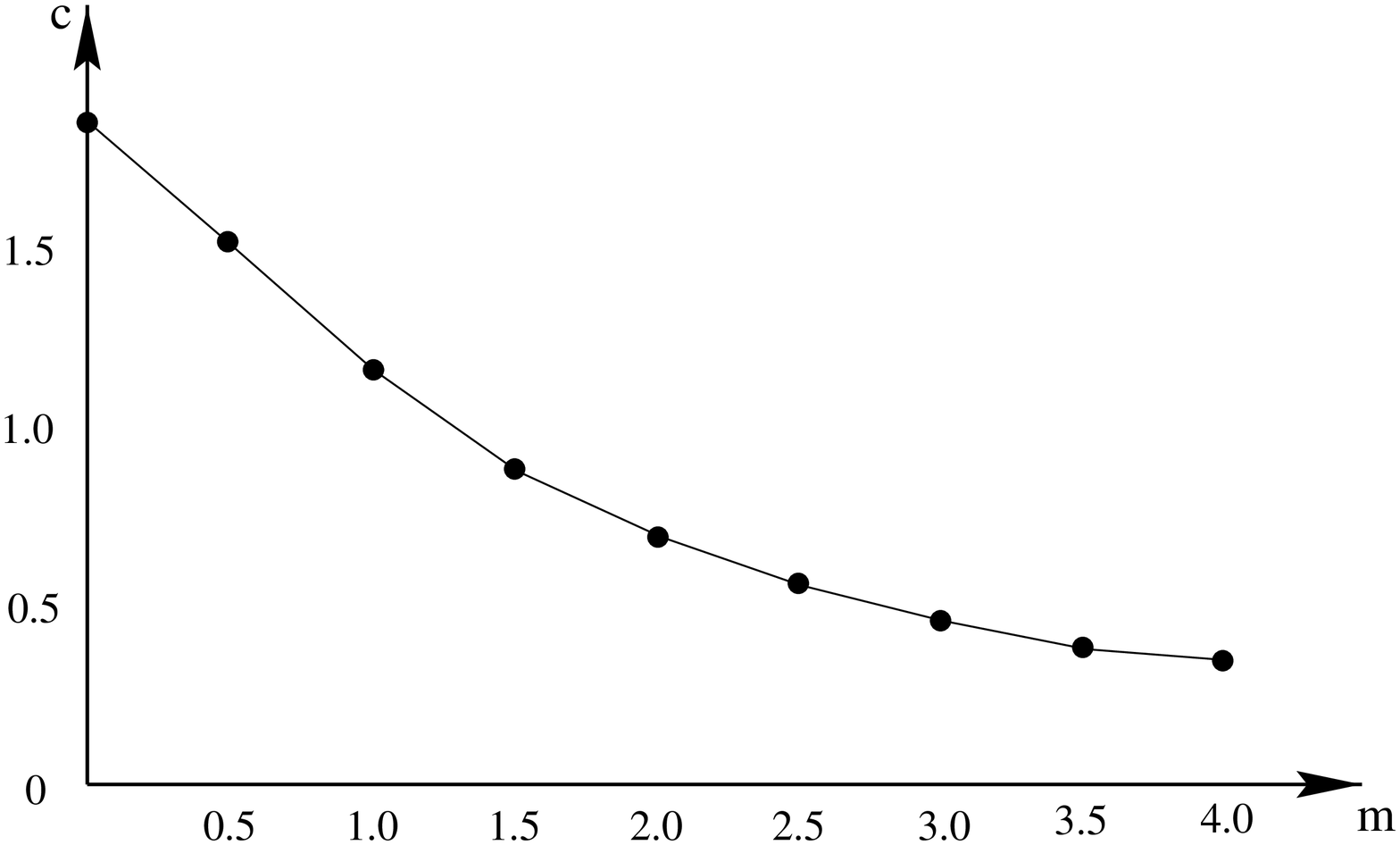}
\caption{Regular solutions of the D7-brane probe equation of motion in the Constable-Myers background. }
\label{quiv8}
\end{center}
\end{figure}

In figure 1b we plot $c$ as a function of $m$ for the regular solutions.  The
numerical evidence suggests that there is a non-zero condensate in the limit
$m\rightarrow 0$. This corresponds to spontaneous breaking of the $U(1)$
chiral symmetry by the quark condensate $c$. 

The condensate seems to screen the probe from the naked singularity. This is
reminiscent of the enhan\c con mechanism \cite{Johnson:1999qt} found in ${\cal
  N}=2$ gravity duals, where the singularity of the geometry is screened from
the physics of a D3-brane probe.  It is possible we are seeing hints of
something similar, if more complicated, here, although at this stage we can
not see how to remove the singularity.

\section{Large $N$ Goldstone boson ($\eta'$)}

Since there is chiral symmetry breaking via a condensate in the
$m\rightarrow 0$ limit,  we expect there to be a Goldstone boson
in the meson spectrum. Such a Goldstone mode must exist as a
solution to the DBI equation of motion, as the following
holographic version of the Goldstone theorem shows. Assume a
D7-embedding with $w_5 = 0$ and $w_6 \sim c/\rho^2$
asymptotically.  A small $U(1)$ rotation $\exp(i\epsilon)$ of $w_5
+i w_6$ generates a solution which is a normalizable small
fluctuation about this background, with $w_6$ unchanged (to order
$(\epsilon^2)$) and $w_5 = \epsilon c/\rho^2$. Thus a small
fluctuation with $w_6$ unchanged and $w_5 = \epsilon
\frac{c}{\rho^2} \sin( k \cdot x)$ is a normalizable solution of
the {\it linearized} equations of motion provided $k^2 = 0$.  Thus
there must be a massless meson associated with the $w_5$
fluctuations. Note that if the embedding were asymptotically $w_6
\sim m + c(m)/\rho^2$ for non-zero $m$,   a $U(1)$ rotation of
$w_5 + i w_6$ would still generate another solution. However this
solution is no longer a {\it normalizable} small fluctuation about
the original embedding.  Thus we no longer expect a massless
meson,  which reflects the explicit symmetry breaking by the
non-zero quark mass $m$.

The existence of a Goldstone boson can be verified numerically. In
the vacuum solutions discussed where $w_6$ has a background value,
fluctuations in $w_5$ should contain the Goldstone mode. Therefore
we solve the linearized equation of motion numerically for
normalizable small fluctuations of the form $w_5 = f(\rho)\sin( k
\cdot x)$, with $x$ the four Minkowski coordinates and $k^2=-M^2$. For the
regular solutions, the meson mass is a function of quark mass
which is plotted in figure 3. The meson mass indeed falls to zero
as the quark mass is taken to zero, providing further evidence of
chiral symmetry breaking. Furthermore, at small $m$ the meson mass
scales like $\sqrt{m}$. This scaling is well known in QCD, and can
be derived from a chiral Lagrangian.

For comparison it is interesting to study $w_6$ fluctuations as
well. The numerical solutions for the $w_6$ fluctuations are also
plotted in figure 3. The  $w_6$ fluctuations have a mass gap, as
expected since they are transverse to the vacuum manifold.
\vspace{-.1cm}
\begin{figure}[!h]
\begin{center}
\includegraphics[height=6cm,clip=true,keepaspectratio=true]{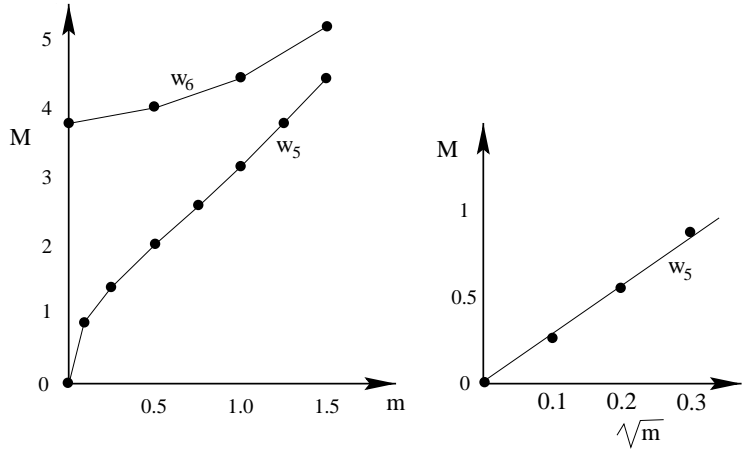}
\vspace{-.3cm}
\caption{A plot of the $w_5$ and $w_6$
meson mass vs quark mass $m$ associated with the fluctuations about
the regular solutions of the equation of motion for the
Constable-Myers flow. The Goldstone mass is also plotted vs $\sqrt{m}$
with a linear fit.}
\label{quiv9}
\end{center}
\end{figure}

\vspace{-.2cm}
Note that the $U(1)$ chiral symmetry is non-anomalous only in the
limit $N\rightarrow\infty$, which is also the limit where
classical supergravity and Dirac-Born-Infeld theory are reliable.
The Goldstone boson discussed above is analogous to the
$\eta^{\prime}$ in QCD, which becomes a Goldstone boson in the
large $N$ limit \cite{Wittenetaprime}. We expect that finite $N$
stringy effects will give the $\eta^{\prime}$ a non-zero mass. A
presently unsolved challenge is to obtain a background
corresponding to spontaneously broken $U(N_f) \times U(N_f)$
chiral symmetry for $N_f>1$.  Embedding $N_f$ D7-branes in the
Constable-Myers background gives a diagonal $U(N_f)$ and an axial
$U(1)$.  There is no larger chiral symmetry because of couplings
$\tilde\psi X \psi$ which involve adjoint scalars $X$.  Note that
the axial $U(1)$ symmetry also acts on $X$.

It is nevertheless interesting to make a rough comparison between
our $\eta'$ and the QCD pions: In a large $N$ QCD model with 
more than one degenerate quark flavor, there would be $N_f^2$
degenerate Goldstone bosons including the usual pions as well as
$\eta'$. In standard QCD the bare up or down quark
mass is roughly $0.01 \Lambda$ and the pion mass of order $0.5
\Lambda$ (it is of course hard to know precisely what value one
should pick for the strong-coupling scale $\Lambda$). If we assume
that $\Lambda \simeq b = 1$ then for this quark mass of $0.01 \Lambda$ 
we find $m_\pi \simeq 0.25 \Lambda$. The gravity dual is correctly
predicting the pion mass at the level of a factor of two. Of
course we cannot expect a perfect match, given the large $N$
limit as well as the additional degrees of freedom which are present 
in the deformed ${\cal N}=4$ theory.

In conclusion,  we have demonstrated that holography
captures one of the most important qualitative features of a 
confining QCD-like theory, namely chiral symmetry breaking by a quark
condensate.  So far we have resorted to numerics for solving the relevant
equations of motion. A natural next step is to understand the main
features of the solutions analytically. Moreover it would be
interesting to see if chiral symmetry breaking may be obtained in
even more realistic backgrounds.  This would
open up the possibility of studying the light
mesonic sector of QCD using the new techniques presented here.

{\bf Acknowledgement} 
We are very grateful to R.~Brower, N.~Constable, A.~Hanany,
C.~N\'{u}\~{n}ez, M.~Petrini and
N.~Prezas for enlightening discussions.
The research of J.E., Z.G.~and I.K.~is supported by DFG
within the `Emmy Noether' programme.
J.B.~is grateful for support through
the A.~von Humboldt Foundation and
N.E. for the support of a PPARC Advanced Research Fellowship.


\vspace{-.3cm}
\begin{thebibliography}{77}

\bibitem{maldacena}
J.M. Maldacena, Adv. Theor. Math. Phys. {\bf 2} (1998) 231
[arXiv:hep-th/9711200].


\bibitem{Witten}
E.~Witten,
Adv.\ Theor.\ Math.\ Phys.\  {\bf 2} (1998) 505
[arXiv:hep-th/9803131].

\bibitem{Gubser:1999pk}
S.~S.~Gubser,
arXiv:hep-th/9902155.

\bibitem{Babington:2002ci}
J.~Babington, D.~E.~Crooks and N.~Evans,
JHEP {\bf 0302} (2003) 024
[arXiv:hep-th/0207076]; 
J.~Babington, D.~E.~Crooks and N.~Evans,
Phys.\ Rev.\ D {\bf 67} (2003) 066007
[arXiv:hep-th/0210068];
N.~Evans, ``QCD-like Gauge Dynamics From Gravity Duals,'' same proceedings.


\bibitem{CM}
N.~R.~Constable and R.~C.~Myers,
JHEP {\bf 9911} (1999) 020
[arXiv:hep-th/9905081];



\bibitem{Csaki}
C.~Csaki, H.~Ooguri, Y.~Oz and J.~Terning,
JHEP {\bf 9901}, 017 (1999)
[arXiv:hep-th/9806021];
R.~C.~Brower, S.~D.~Mathur and C.~I.~Tan,
Nucl.\ Phys.\ B {\bf 587} (2000) 249
[arXiv:hep-th/0003115].

\bibitem{Teper}
M.~Teper,
arXiv:hep-ph/0203203;
C.~J.~Morningstar and M.~J.~Peardon,
Phys.\ Rev.\ D {\bf 60}, 034509 (1999)
[arXiv:hep-lat/9901004];
%
D.~E.~Crooks and N.~Evans,
arXiv:hep-th/0302098.

\bibitem{KarchRandall}
A.~Karch and L.~Randall,
JHEP {\bf 0106} (2001) 063
[arXiv:hep-th/0105132].

\bibitem{DFO} O.~DeWolfe, D.~Z.~Freedman and H.~Ooguri,
Phys.\ Rev.\ D {\bf 66} (2002) 025009
[arXiv:hep-th/0111135].

\bibitem{CEGK1}
N.~R.~Constable, J.~Erdmenger, Z.~Guralnik and I.~Kirsch,
arXiv:hep-th/0211222.


\bibitem{KarchKatz}
A.~Karch and E.~Katz,
JHEP {\bf 0206} (2002) 043
[arXiv:hep-th/0205236];


\bibitem{MateosMyers}
M.~Kruczenski, D.~Mateos, R.~C.~Myers and D.~J.~Winters,
arXiv:hep-th/0304032.



\bibitem{Sakai:2003wu}
T.~Sakai and J.~Sonnenschein,
JHEP {\bf 0309} (2003) 047
[arXiv:hep-th/0305049].


\bibitem{Babington:2003vm}
J.~Babington, J.~Erdmenger, N.~Evans, Z.~Guralnik and I.~Kirsch,
arXiv:hep-th/0306018.


\bibitem{Gubser:2000nd}
S.~S.~Gubser,
Adv.\ Theor.\ Math.\ Phys.\  {\bf 4} (2002) 679
[arXiv:hep-th/0002160].


\bibitem{Wittenetaprime} E.~Witten,
Nucl.\ Phys.\ B {\bf 156} (1979) 269.




\bibitem{Wang:2003yc}
X.~J.~Wang and S.~Hu,
JHEP {\bf 0309} (2003) 017
[arXiv:hep-th/0307218].

\bibitem{Ouyang}
P.~Ouyang,
arXiv:hep-th/0311084.


\bibitem{Nunez:2003cf}
C.~Nunez, A.~Paredes and A.~V.~Ramallo,
arXiv:hep-th/0311201.

\bibitem{Kruczenski:2003uq}
M.~Kruczenski, D.~Mateos, R.~C.~Myers and D.~J.~Winters,
arXiv:hep-th/0311270.


\bibitem{Hong:2003jm}
S.~Hong, S.~Yoon and M.~J.~Strassler,
arXiv:hep-th/0312071.


\bibitem{Johnson:1999qt}
C.~V.~Johnson, A.~W.~Peet and J.~Polchinski,
Phys.\ Rev.\ D {\bf 61} (2000) 086001
[arXiv:hep-th/9911161].



\end{thebibliography}
\end{document}